\documentclass[amsmath,aps,fleqn]{article}
\pdfoutput=1

\usepackage{jcappub}
\usepackage{amssymb}
\usepackage{amsmath}
\usepackage{epsfig}
\usepackage{subfigure}
\usepackage{mathrsfs}
\usepackage{longtable}
\usepackage{aas_macros}

\newcommand{\be}{\begin{eqnarray}}
\newcommand{\ee}{\end{eqnarray}}
\newcommand{\bea}{\begin{eqnarray}}
\newcommand{\eea}{\end{eqnarray}}

\newcommand{\geff}{\ensuremath{G_\mathrm{eff}}}

\title{Ambiguous Tests of General Relativity on Cosmological Scales}
\author[a,b,c,1]{Joe Zuntz}
\note{jaz@astro.ox.ac.uk}
\author[a,2]{Tessa Baker}
\note{tessa.baker@astro.ox.ac.uk}
\author[a,3]{Pedro G. Ferreira}
\note{p.ferreira1@physics.ox.ac.uk}
\note{skordis@nottingham.ox.ac.uk}
\author[d,4]{Constantinos Skordis}
\affiliation[a]{Astrophysics, University of Oxford, DWB, Keble Road, Oxford, OX1 3RH, UK}
\affiliation[b]{Department of Physics \& Astronomy, University College London, WC1E 6BT, UK}
\affiliation[c]{Oxford Martin School, University of Oxford, 34 Broad St, Oxford OX1 3BD, UK}
\affiliation[d]{School of Physics and Astronomy, University of Nottingham, University Park, Nottingham,  NG7 2RD,UK}

\abstract{
There are a number of approaches to testing General Relativity (GR) on linear scales using parameterized frameworks for modifying cosmological perturbation theory. It is sometimes assumed that the details of any given parameterization are unimportant if one uses it as a diagnostic for deviations from GR. In this brief report we argue that this is not necessarily so.  First we show that adopting alternative combinations of modifications to the field equations significantly changes the constraints that one obtains.  In addition, we show that using a parameterization with insufficient freedom significantly tightens the apparent theoretical constraints.  Fundamentally we argue that it is almost never appropriate to consider modifications to the perturbed Einstein equations as being constraints on the effective gravitational constant, for example,  in the same sense that solar system constraints are.  The only consistent modifications are either those that grant near-total freedom, as in decomposition methods, or ones which map directly to a particular part of theory space.}
\begin{document}

\maketitle

\section{Introduction}

\label{Intro}
The availability of high-precision cosmological data has made it possible to test General Relativity (GR) on cosmological scales.  In particular, in the absence of a conclusive theoretical explanation for the dark energy phenomenon, modifications and extensions to that venerable theory remain a viable possibility.

Since the process of comparing a new theory of gravity to cosmological data can be complicated, any test statistic that can be used to quickly reject a theory is a useful one.  In this spirit, inspired by the Parameterized Post-Newtonian (PPN) formalism \cite{Will_living_review_2006,Will1971, Thorne_Will,Will_Nordvedt_1972}, a similar approach has been developed, which describes the deviations of a theory from the standard evolution of a perturbed GR universe \cite{AmendolaKunzSapone,AminBlandfordWagoner,ZhangEtAl,Pogosian_parameterization,Daniel2010,Hu_Sawicki,Daniel_Linder,Hojjati_Pogosian,Song2010,Bertschinger2006,Zhao2010,bean,Caldwell,skordis,skordisFerreira,baker,isitgr1,isitgr2}.  Unlike the PPN case, however, such a parameterization is not unique; on cosmological scales there is freedom to modify evolution over the whole of cosmic time and over a range of scales. One is therefore led to constrain functions of space and time rather than just a handful of numbers in the weak-field metric.
There are then two levels of structure that must be imposed on the parameterized field equations:
\begin{description}
\item[Level 1:] One must choose the way in which free functions are used to extend the field equations. For example, one could use free functions to rescale terms that exist in GR, or allow new terms to be present.
\item[Level 2:] When implementing the parameterized field equations in numerical codes one is forced to choose a sensible ansatz for the time- and scale-dependence of the free functions. Typically these ansatzes are motivated by the observation that the dark energy-like sector only dominates the energy density of the universe at late cosmological times. See \cite{MG_report} for an enumeration of such ansatzes.
\end{description}
The choice usually made at level 1 is to introduce two free functions directly into the modified Poisson equation and the  `slip relation' (the transverse, traceless component of the field equations), as these are the expressions relevant to observables such as weak lensing of galaxies and measures of structure growth \cite{bean,Caldwell} . One of these free functions acts as an effective rescaling of Newton's gravitational constant, whilst the other is defined as the ratio of the two potentials that describe the perturbed metric in the Conformal Newtonian gauge. We will refer to this approach as `phenomenologically-based'. It is \textit{not} the case that particular choices of the two free functions are designed to reproduce the field equations of specific modified gravity theories (except in a handful of special cases). Rather, in the phenomenological approach, the free functions should be considered as indicator flags for non-GR behaviour.

A disadvantage of phenomenological-type parameterizations is their tendency to obscure which regions of theory space they correspond to, because there is no direct mapping between the free functions and the parameters of a specific model. It is therefore difficult to translate constraints on the two free functions into constraints on a given theory of modified gravity.

We have advocated an alternative approach: directly specifying quantities needed to describe a $4\times 4$ tensor of scalar modifications to the linearly perturbed Einstein equations \cite{skordis,skordisFerreira,baker}. We can then derive the corresponding Poisson equation and slip relation, which will contain components of this new tensor.This has been dubbed the `Parameterized Post-Friedmann' approach (PPF), and it leads to a level-1 structure (as classified above) which is different to that of the phenomenological approach.

Arguments can be made in favour of both the phenomenologically-based and PPF approaches. For large classes of theories the two strategies become equivalent at intermediate distance scales where the time derivatives of perturbations can be neglected in comparison with their spatial derivatives. However, this approximation is not applicable on scales comparable to the horizon size. 

Whilst such large scales cannot be probed directly with galaxy surveys, they are nonetheless important for accurate calculation of the Integrated Sachs-Wolfe (ISW) effect and large-scale matter power spectrum. These quantities are frequently computed using Einstein-Boltzmann solvers such as {\sc Camb} \cite{camb} or {\sc Class} \cite{class} which evolve perturbations through horizon crossing. 

The purpose of this brief report is to illustrate that the choice made for the level-1 structure of a parameterization has a significant influence on the constraints obtained on deviations from GR -- a caveat that is commonly forgotten. 

We demonstrate the differences between two possible parameterizations by generating some of the  simpler perturbative observables - the CMB power spectra, in particular the ISW effect, matter power spectra, and the growth function $f(z)$. In \textsection\ref{ansatzes} we further show that the choice made for the level-2 structure has an equally important influence on the constraints obtained.

\section{Theory}
\label{theory section}

In this section we describe the two parameterizations that we apply in this paper.  Our goal here is not to advocate for either of them, but rather to highlight the fact that they lead to different results. Parameterization A described below is derived as a special case of the general PPF form detailed in \cite{baker_etal_2012}. However,  for the purposes of this paper we can ignore its origin and simply regard the resulting field equations as another example of a phenomenological-type parameterization -- one with different choices made for the level-1 structure.

\subsection{Parameterization A}
In \cite{skordis} a parameterization was proposed by writing the modifications as an additional tensor to the Einstein equations of GR:
\begin{equation}
\label{EFE}
G_{\mu\nu}=8\pi G_0 a^2 T^M_{\mu\nu}+ a^2 U_{\mu\nu}
\end{equation}
where the stress-energy tensor $T^M_{\mu\nu}$ contains all the standard cosmologically-relevant fluids and the tensor $U_{\mu\nu}$ may contain metric, matter and additional field degrees of freedom coming from a theory of gravity.
In writing (\ref{EFE}) it is assumed that all known matter fields which are part of $T^M_{\mu\nu}$ couple to the same metric $g_{\mu\nu}$, and that $G_{\mu\nu}$ is the Einstein tensor of that same metric. 
This ensures that the stress-energy tensor of matter obeys its usual conservation equations, and hence  $U_{\mu\nu}$ is separately conserved.

The formalism proceeds by parameterizing around the linearly perturbed version of equation (\ref{EFE}). 
To enable a direct comparison with parameterization B below we will specialise to the case of purely metric theories, that is, those for which the action is constructed from curvature invariants (e.g. $f(R)$ gravity) or non-local invariants (e.g. \cite{DeserWoodard}).

The most general perturbations of $U_{\mu\nu}$ that are allowed in a second-order metric-only theory are as follows \cite{skordis,skordisFerreira,baker, baker_etal_2012}:
\begin{eqnarray}
 \label{2nd_order_Us}
-a^2\delta U^0_0 &=& k^2 A_0\hat\Phi\nonumber\\  
-a^2 \delta U_i^0 &=&k B_0 \hat\Phi\nonumber \\
a^2\delta U_i^i &=&k^2 C_0 \hat\Phi + k C_1 \dot{\hat\Phi} \nonumber\\
a^2\delta U_j^i &=&D_0 \hat\Phi+\frac{D_1}{k} \dot{\hat\Phi}
\end{eqnarray} 
where dots denote derivatives with respect to conformal time $\eta$ and $k$ is the Fourier wavenumber.
The gauge-invariant metric perturbation $\hat\Phi$ reduces to the curvature perturbation $\Phi$ in the Conformal Newtonian (CN) gauge. In our conventions the CN gauge is defined by:
 \begin{equation}
 ds^2 = -a^2(1 + 2 \Psi) d\eta^2 + a^2 (1 - 2\Phi) d\vec{x}^2
 \end{equation}
Further below we introduce a second gauge-invariant metric perturbation, $\hat{\Psi}$
which reduces to $\Psi$ in the CN gauge.

The coefficients $A_0\ldots D_1$ appearing above are functions of background quantities, $A_0=A_0 (k, \eta)$ etc., and the factors of $k$ are chosen such that $A_0\ldots D_1$ are dimensionless. However, these functions are not all independent. Perturbations of the Bianchi identity  $U^{\mu}_{\nu;\mu}=0$ yield a set of additional constraint equations which can be used to reduce the six free functions in equations (\ref{2nd_order_Us}) down to just two. One of these free functions is defined to be:
\begin{equation}
\frac{D_1}{k}=\frac{\tilde g}{\cal H},\quad\;\mathrm{where}\quad\; \tilde{g}=-\frac{1}{2}\left(A_0+3\frac{\cal H}{k}B_0\right)
\end{equation}
The modified Poisson equation can then be written (in Fourier space):
\begin{eqnarray}
-k^2\hat\Phi&=&4\pi \frac{G_0}{1-\tilde{g}} a^2 \rho\Delta  \equiv 4\pi G_\mathrm{eff}a^2 \rho\Delta   \label{Poisson}
\end{eqnarray}
with $G_0$ denoting the canonical value of Newton's gravitational constant as measured by a Cavendish experiment on the Earth, and where the gauge-invariant comoving density perturbation $\rho\Delta$ is a summation over all conventional cosmologically-relevant fluids. The combination $G_\mathrm{eff}=G_0\,(1-\tilde{g})^{-1}$ plays the role of a modified Newton's constant. 

We choose the second free function to be $D_0$, which we will hereafter relabel as $\zeta=\zeta (k, \eta)$ to distinguish it from its appearance in the more general format of equations (\ref{2nd_order_Us}). 
$\zeta$ appears in the `slip' relation between the potentials $\hat\Phi$ and $\hat\Psi$ (which are equal to one another in GR): 
\begin{equation}
\label{slip}
\hat\Phi-\hat\Psi=8\pi G_0 a^2 (\rho+P)\Sigma+\zeta\hat\Phi+\frac{\tilde g}{\cal H}\dot{\hat\Phi}
\end{equation}
The anisotropic stress of conventional matter $\Sigma$ is negligible after the radiation-dominated era.

The key point here is that in parameterization A the slip relation and the modified Newton's constant are not independent, as the function $\tilde{g}$ appears in both. This special case of the PPF formalism does not capture the behaviour of many modified gravity theories, because we have not allowed for additional degrees of freedom to appear \cite{baker, baker_etal_2012}. It is, however, directly comparable to a phenomenological-type parameterization. Hence we will regard eqns.(\ref{Poisson}) and (\ref{slip}) simply as possible alternatives to eqns.(\ref{phenom_Poisson}) and (\ref{phenom_slip}).

\subsection{Parameterization B}

If one steps back and looks at the structure of the evolution equations for cosmological perturbations, 
one finds that it is enough to work with only two of the four Einstein field equations: the Newton-Poisson equation and the slip equation. From a practical point of view, the only observable modifications to
gravity (at least at the perturbative level) will be modifications to the these equations, which we can
write in the following form
\begin{eqnarray} 
-k^2\Phi-4\pi G_0 a^2\rho\Delta &=&F_1  \label{phenom_Poisson0}\\
(\Psi-\Phi)+8\pi G_0 a^2 (\rho+P)\Sigma&=& F_2\label{phenom_slip0}  
\end{eqnarray}
where $F_1$ and $F_2$ are arbitrary functions of time {\it and} space. A simple ansatz (from the
dimensional point of view) is that $F_1=\alpha k^2\Phi$ and $F_2=-\zeta\Phi$. Reorganizing the equation we find that they can be rewritten as
\begin{eqnarray} 
-k^2\Phi&=&4\pi G_\mathrm{eff} a^2\rho\Delta  \label{phenom_Poisson}\\
\Psi&=&-8\pi G_0 a^2 (\rho+P)\Sigma+(1-\zeta)\Phi \label{phenom_slip}  
\end{eqnarray}
Parameterization B is defined in the CN gauge, so that $\Phi$ and $\Psi$ replace $\hat\Phi$ and $\hat\Psi$ in eqns.(\ref{Poisson}) and (\ref{slip}). This formulation is equivalent up to small corrections to that used in \cite{bean}; their anisotropic stress is related to ours by $k^2\Sigma=\sigma$. This parameterization suggests that, once the anisotropic stress has become negligible, a theory of modified gravity could potentially modify one of equations (\ref{phenom_Poisson}) or (\ref{phenom_slip}) whilst leaving the other unchanged.  This behaviour does not arise \textit{analytically} from theories below third order \cite{baker}; but given enough freedom in the functional ansatz the \textit{numerical} behaviour that occurs in any theory can be realised with this parameterization. (We note that a handful of parameterizations that are instead optimized for weak lensing analysis have also been suggested \cite{Song2010,Zhao2010,Daniel_Linder}). 

To demonstrate the differences that the alternative parameterization types can generate,  we will use the same ansatz for the free functions $\geff$ and $\zeta$ in both parameterizations A and B. Since we are focusing on modifications to gravity associated with dark energy, an appropriate parameterization is an expansion in $\Omega_\Lambda$.  We therefore use a polynomial expansion up to third order in $\Omega_\Lambda$.
\section{Spectra}
We have modified a version of the Boltzmann code {\sc Camb} to use the new equations specified in section \ref{theory section}.  We use variants which include both  parameterizations A and B.  In this section we show and analyze CMB and LSS spectra generated by these models.  We will show that the two variants, even when set up with equivalently parameterized functions, generate considerably different spectra. 

\begin{figure}[htbp]
\begin{center}
\includegraphics[width=8cm]{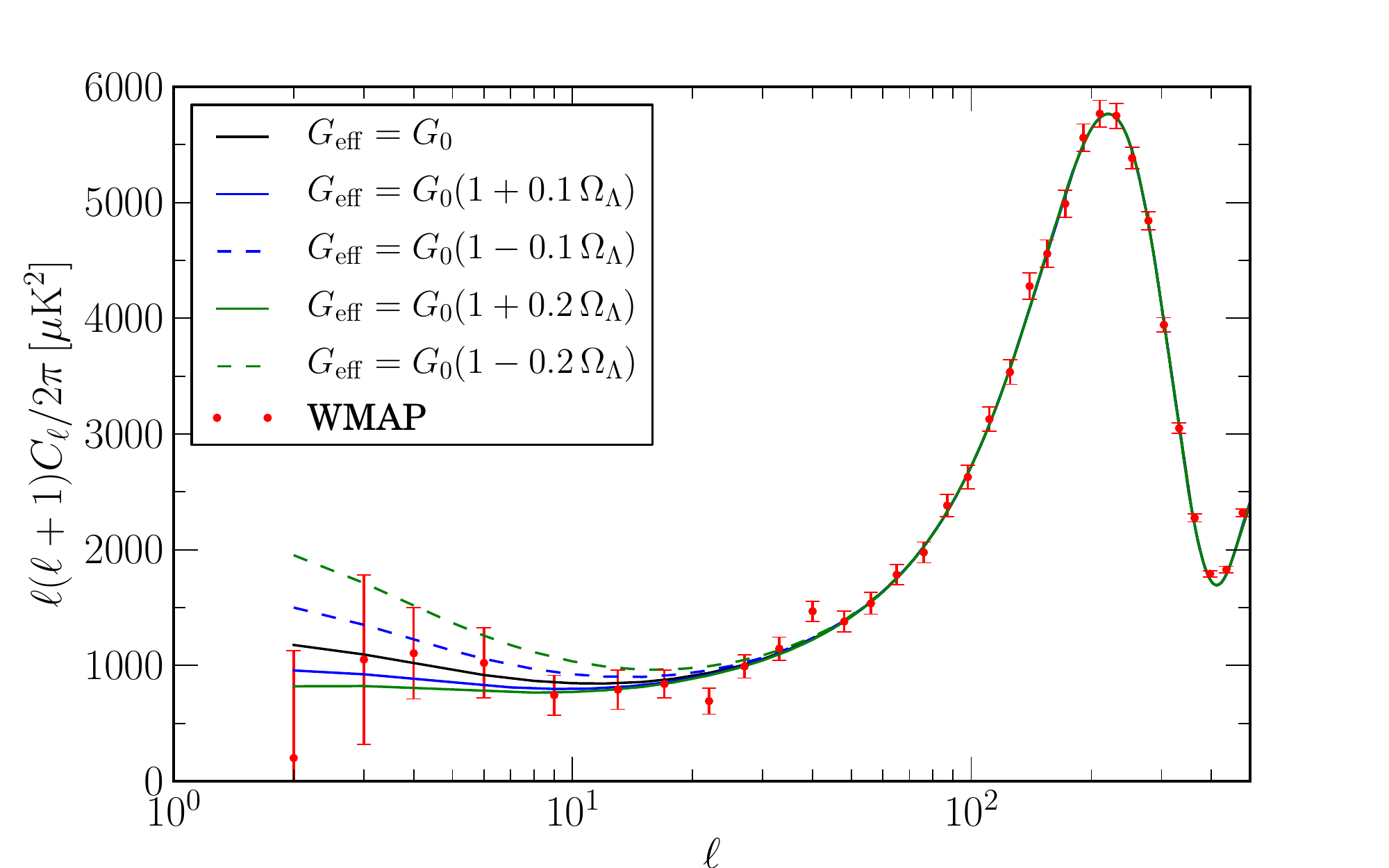}
\caption{CMB power spectra for varying the linear term of the Taylor expansion in $\Omega_\Lambda$ of the $G_{\mathrm{eff}}$ appearing in the Poisson equation, as defined in equation (\ref{Poisson}).}
\label{linear geff plot}
\end{center}
\end{figure}

Figure \ref{linear geff plot} shows CMB spectra from parameterization A with variations to the linear term.  Because we have modified the late-time behaviour of gravity whilst leaving the era before dark energy domination unchanged, the small-scale CMB (which depends mainly on what happens at recombination) is unaffected, but the large-scale features caused by the ISW effect vary.  It can be seen that strengthening gravity by a small amount decreases the power spectrum of the ISW effect by counteracting the suppressive effect of $\Lambda$ on structure formation, whereas weakening $G_{\mathrm{eff}}$ increases the ISW effect.  This difference is significant because the ISW effect is only weakly present observationally.

The behaviour of the spectra for a simple linear variation in $G_{\mathrm{eff}}$ shows the same general trends for model B as those for model A shown in figure \ref{linear geff plot}. The most important difference between the theories, however, appears when the two free functions interact to allow cancellations. This is discussed in detail below.

Examining the plot of linear variations in $G_{\mathrm{eff}}$ alone, one might be tempted to conclude that the perturbative strength of gravity, as measured by the $G_{\mathrm{eff}}$ parameter, is rather well constrained using just CMB data.  We wish to emphasise in this paper that this is not so.  First, the various ways one can construct a theory mean that direct interpretation of the parameter, except on small scales, is not unique.  Secondly and probably more importantly, the linear variation alone masks a richer phenomenology, which we will discuss more in the next section.

\begin{figure}[htbp]
\begin{center}
\includegraphics[width=8cm]{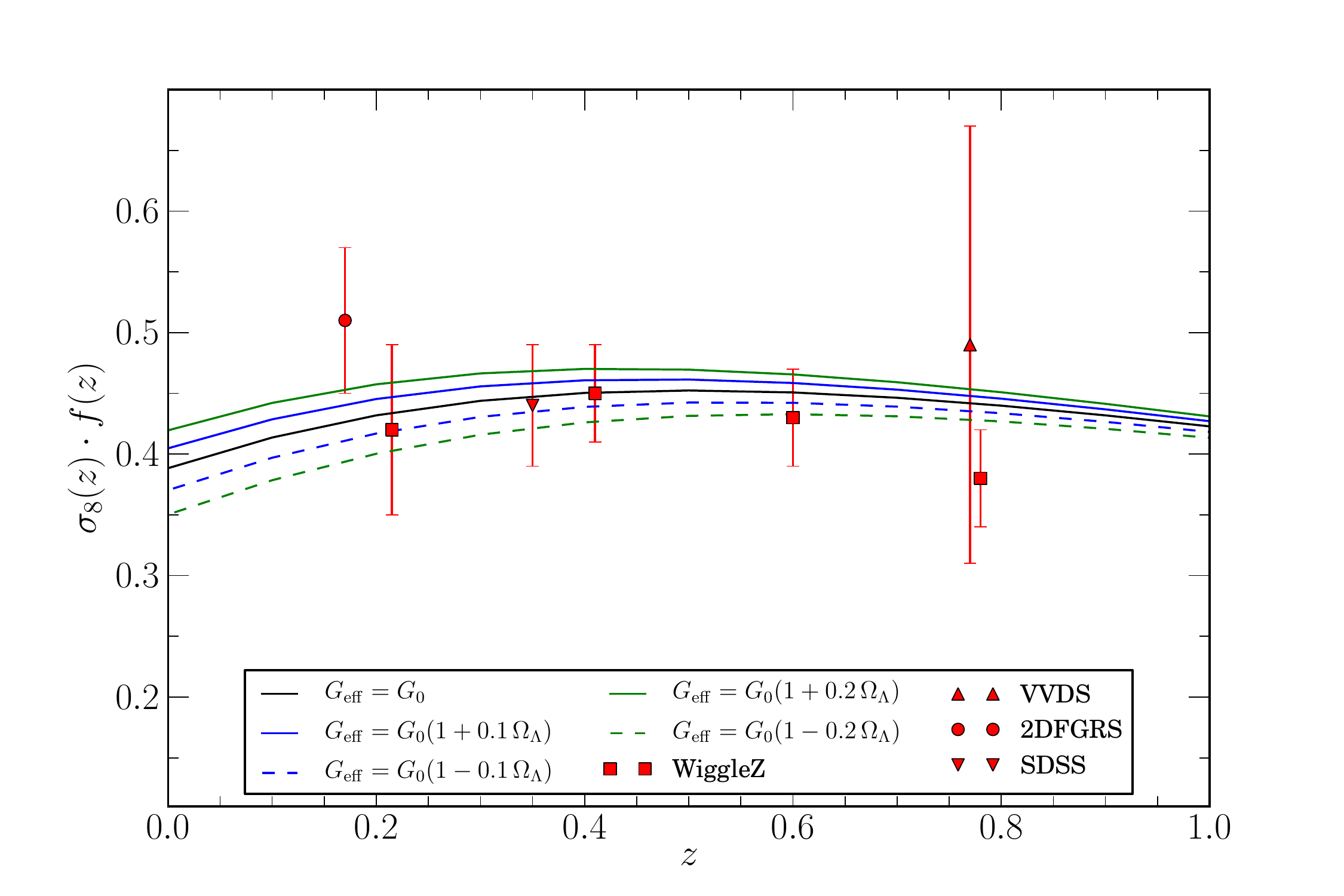}
\caption{The growth function $\sigma_8(z) f(z)$ for the same models shown in figure \ref{linear geff plot}.}
\label{growth function plot}
\end{center}
\end{figure}

Figure \ref{growth function plot} shows the growth function, 
$f(z) = \frac{1}{\cal H}\frac{\dot{(\delta_\rho)}}{\delta_\rho}$
scaled by $\sigma_8$, which is measured by, for example, redshift-space distortion experiments.  Results from the recent WiggleZ survey \cite{wigglez} and other data \cite{Percival2DF,TegmarkSDSS,GuzzoVVDS,SongPercival} are also shown.  These results provide weaker constraints than the ISW, but near-future experiments should reach sensitivity levels good enough for precision tests of late-time gravity.

We do not display the matter power spectra because they are very mildly affected by the modifications of gravity, except on very large scales far beyond the reach of galaxy surveys.

\begin{figure}[htbp]
\begin{center}
\includegraphics[width=8cm]{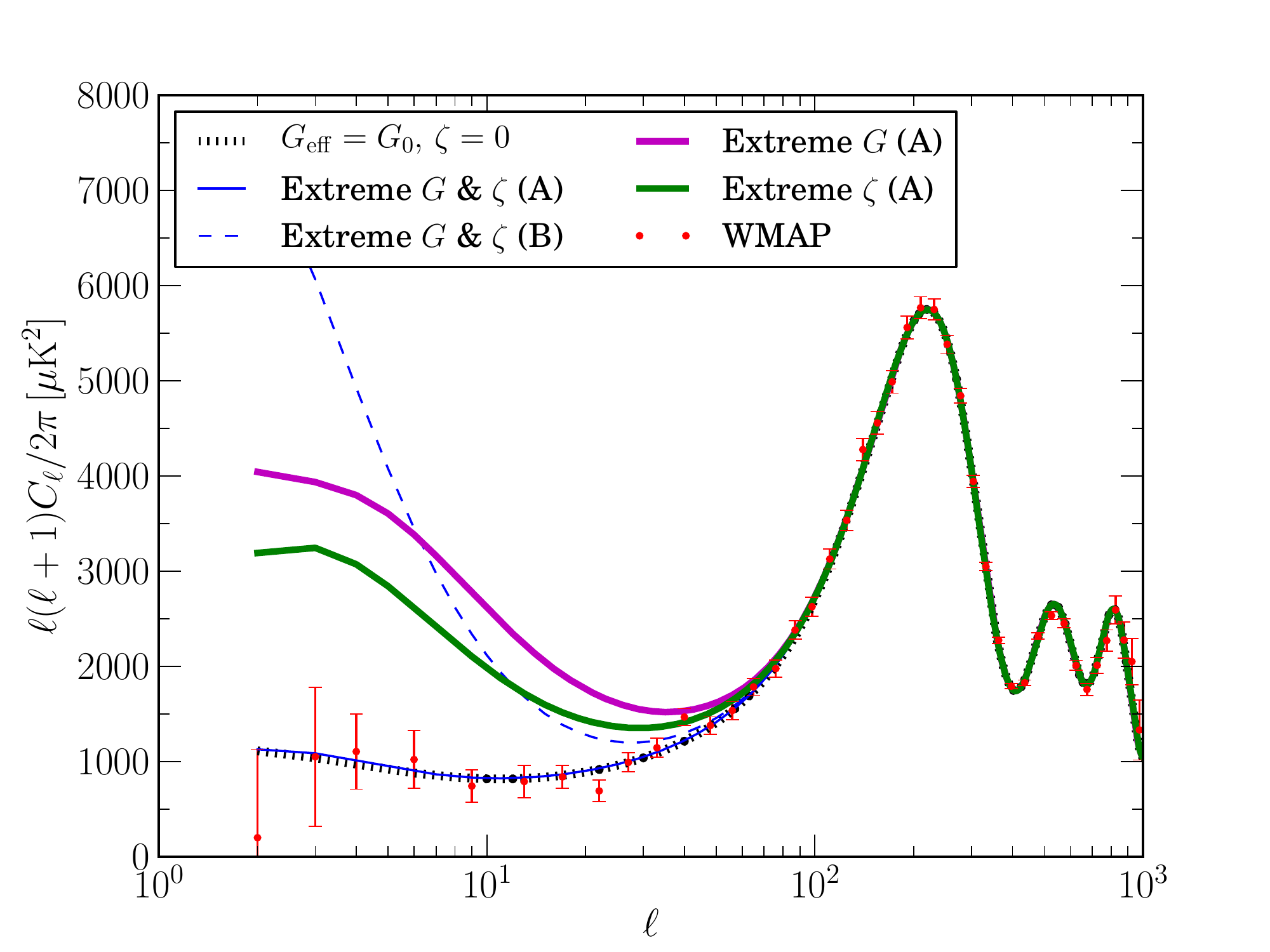}
\caption{A demonstration of how extreme changes to the effective gravitational constant can be permitted by the CMB when they can be counteracted by a significant gravitational slip (see text for form of extreme curves).  The cancellation is absent for these parameters in model B.  This delicate difference underscores the need for a rigorous treatment of modified functions.}
\label{extreme plot}
\end{center}
\end{figure}

Figures \ref{linear geff plot} and \ref{growth function plot} demonstrate that relatively small modifications of about 20\% to the effective Newton's constant in the Poisson equation can have effects that are easily detectable at large scales in the CMB.  But it is the combined effects of \geff\  and $\zeta$ that are of most interest.  Figure \ref{extreme plot} demonstrates how this combination can cancel even quite extreme individual effects.  The curves in this plot have \mbox{$G_\mathrm{eff}/G_0 = 1+5.3\,\Omega_\Lambda-1.1\,\Omega_\Lambda^2+1.3\,\Omega_\Lambda^3$}, and \mbox{$\zeta = 4.8\,\Omega_\Lambda - 1.0\, \Omega_\Lambda^2 + 6.6\,\Omega_\Lambda^3$}. The other cosmological parameters are within normal ranges, though not identical to those in figure \ref{linear geff plot}.  There is a near-total cancellation of the ISW effect in model A, but model B with the same parameters is completely ruled out. There do exist alternative parameter combinations in model B with the same cancellation, but the fact that they are very different from those in model A is what we want to highlight here.

We are not, of course, advocating these models as actual cosmologies - they simply demonstrate that a vast region of parameter space can be consistent with CMB data.   Since other upcoming data sets like lensing and redshift-space distortions also constrain (different) combinations of the potentials there should be equivalent degeneracies present there.

\begin{figure}[htbp]
\begin{center}
\includegraphics[width=8cm]{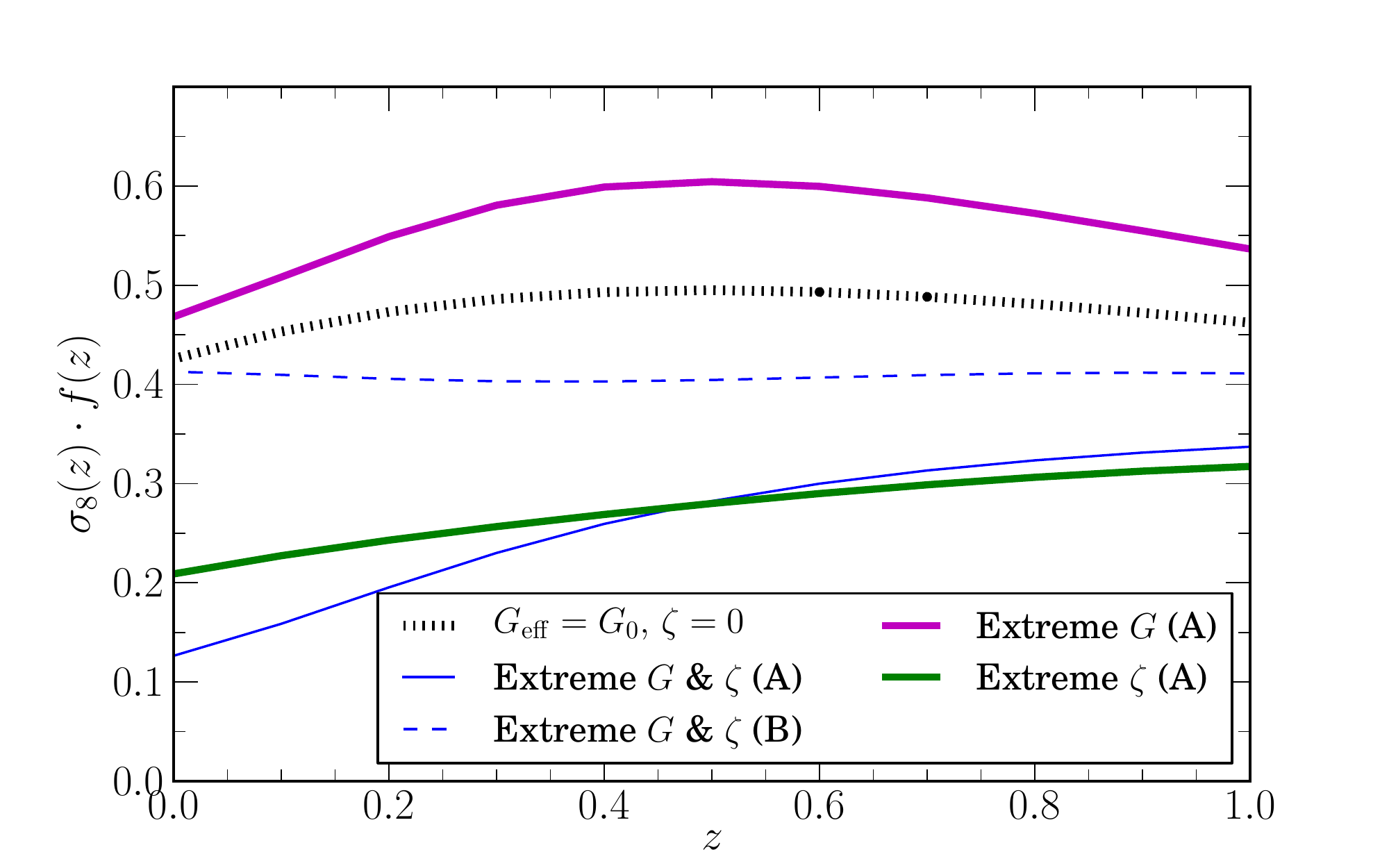}
\caption{The growth rates $f(z)$ corresponding to the CMB spectra shown in figure \ref{extreme plot}.}
\label{extreme growth function plot}
\end{center}
\end{figure}

The importance of other data sets in breaking the degeneracy in the CMB is illustrated in figure \ref{extreme growth function plot}. That plot shows the growth rates $\sigma_8 f(z)$ for the extreme models described above.  The behavior of the extreme case is once again very different between the two parameterizations.

\section{Constraints}
\label{constraints section}

\subsection{Parameter Estimation}
Now that we have demonstrated that the different parameterization approaches can yield considerably different results for power spectra, we can check how these differences go forward into parameter estimation. Note that at this stage we are \emph{not} attempting to find the tightest possible constraints on these parameterizations using all available data.  In particular we are not using data from weak lensing, galaxy-ISW correlations, or growth rates which can provide the strongest constraints.

We are rather intending to demonstrate that the different schemes applied to the same data can produce significantly different constraints, even on quantities we usually regard as rather physical -- such as the the effective Newton's constant controlling the growth of structures under gravity.  We hope to show that headline constraints on the two free functions of parameterized approaches should be taken with a degree of caution.

The plots in this section are generated from Monte-Carlo Markov Chains running parameterizations A (equations (\ref{Poisson}) and (\ref{slip})) and B (equations (\ref{phenom_Poisson}) and (\ref{phenom_slip})).  In each case, as in the previous section, the same ansatz is applied to the free functions $G_\mathrm{eff}(z)$ and $\zeta(z)$.
The same constraining data is used in each case: the 7-year WMAP CMB data \cite{wmap7}, the SDSS DR7 matter power spectrum \cite{sdss-dr7}, a prior $H_0 = 73.8 \pm 2.4$ \cite{riess2011}, the BBN constraint $\Omega_b h^2 = 0.022 \pm 0.002$ \cite{bbn}, and the Union2 Supernova Ia data \cite{Amanullah2010}. \footnote{It is not generally correct to use published supernovae constraints directly when using modified gravitational physics, since the calibration factors applied to them are cosmology-dependent; it is only because we leave our background evolution unchanged from GR that is is possible here.}.
With these data sets the strongest constraining power comes from the ISW effect (or rather lack thereof) in the large-scale CMB temperature power spectrum.

\subsection{Insufficient ansatzes}
\label{ansatzes}

The simplest ambiguity we can consider when constraining the free functions can arise if we use what might be termed an incomplete parameterization -- one where the free functions are not free enough to explore the available parameter space.  One example of this would be our restriction here to functions only of $z$, with no scale-dependence.

\begin{figure}[htbp]
\begin{center}
\includegraphics[width=8cm]{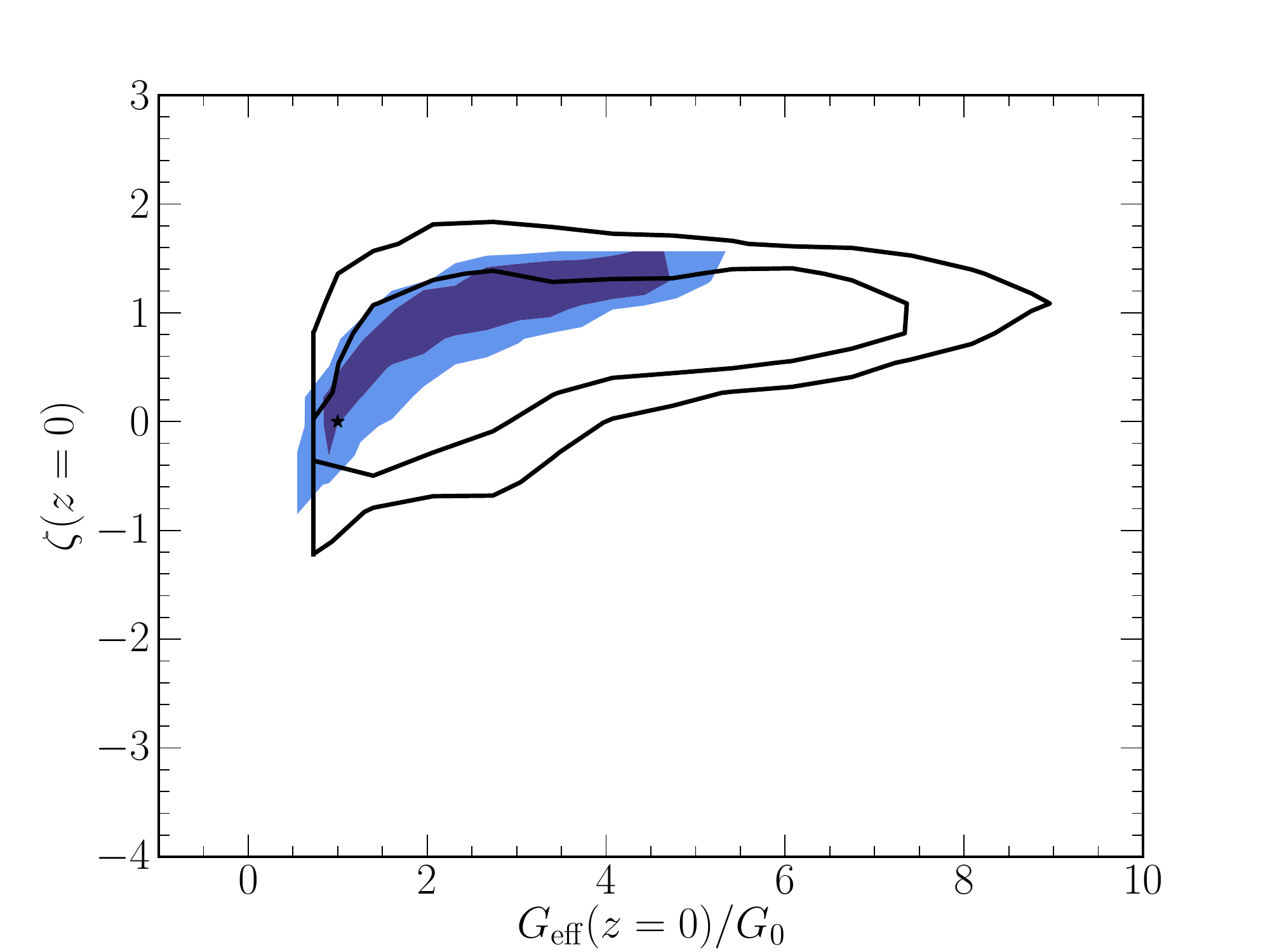}
\caption{Constraints on the free functions, evaluated at redshift zero, in a model restricted to linear evolution in $\Omega_\Lambda$ (filled contours) and one with cubic evolution (line contours).  The apparent constraints on the two parameters, for example, are misleadingly small in the more restrictive parameterization.}
\label{linear constraints}
\end{center}
\end{figure}

 Figure \ref{linear constraints} illustrates another example of an incomplete parameterization. In that figure we show how the constraints on the theory depend on the number of terms in the polynomial expansions of $\geff(\Omega_\Lambda)$ and $\zeta(\Omega_\Lambda)$. The larger contours use cubic models for the evolution, and the much smaller ones use only a linear term in the expansion.  Both curves are for parameterization A, and the contours are for 68\% and 95\% probability mass.

Of course we do not claim that the cubic model is general enough to describe the free functions sufficiently - it is the comparison that we highlight.  Whenever contours are presented for the free functions of some parameterized model one should bear in mind that adding only a little freedom can change the areas of the error ellipses by a large factor. Hence all constraints of this form should be taken with a pinch of salt.

\subsection{Variant parameterization}

\begin{figure}[htbp]
\begin{center}
\includegraphics[width=8cm]{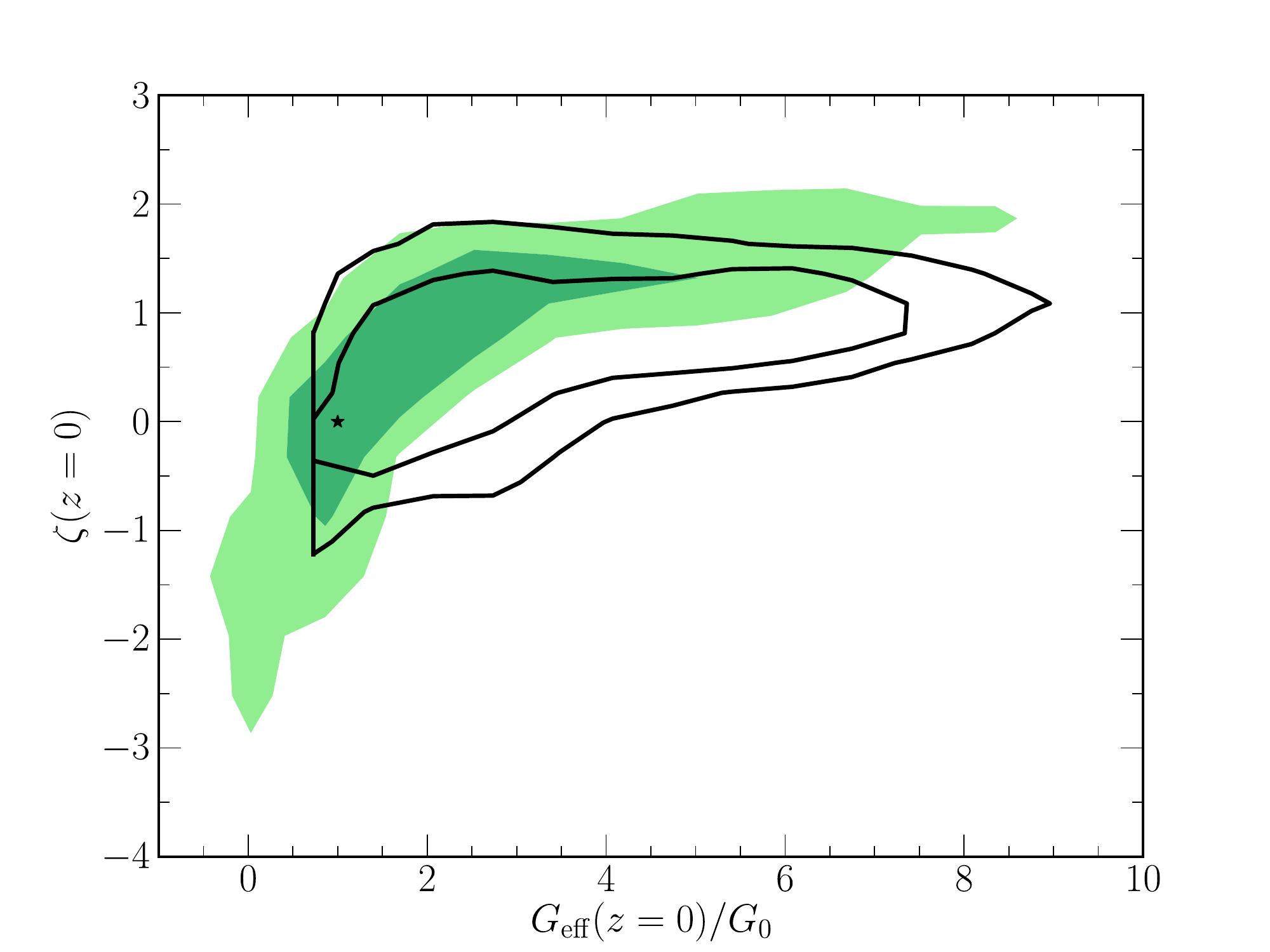}
\caption{Joint constraints on the slip parameter $\zeta$ and $G_\mathrm{eff}$  at $z=0$, for parameterization A (black lines) and B (filled green).  In both cases 68\% and 95\% contours are shown.}
\label{2d histogram}
\end{center}
\end{figure}

The second result of this section is shown in figure \ref{2d histogram}; it demonstrates the parameterization-dependence of the constraints, independent of the ansatz.  Most clear is that the free functions are significantly \emph{less} correlated in model A.  In the model B there is a simple and straightforward impact on the term whose derivative is the ISW source:
\begin{equation}
\Phi + \Psi \propto (\zeta-2) \geff \frac{a^2}{k^2} \rho\Delta
\end{equation}
which yields a multiplicative correlation, whereas case A does not have this simple relation:
\begin{eqnarray}
\Phi &+& \Psi \propto {\cal H}^{-1} \left(\geff-G_0\right)\frac{a^2}{k^2} \rho\dot{\Delta}\\
&&+\left[ (\zeta-3)\geff+ {\cal H}^{-1} \left(1-\frac{G_0}{\geff}\right)\dot{G}_{\mathrm{eff}}+G_0 \right] \frac{a^2}{k^2} \rho\Delta \nonumber
\label{eq isw}
\end{eqnarray}

The edge of the curve for parameterization A in figure \ref{2d histogram} cuts off rather sharply just below $\geff=1$. This is due to the impact of $G_{\mathrm{eff}}$ and $\zeta$ on the growth of matter density perturbations, and the restrictions that they should obey in order to reproduce the observed amount of structure in the universe today. During the matter-dominated epoch density perturbations on subhorizon scales grow as \mbox{$\delta\propto a^{p_i}$}, where in parameterization A \cite{baker2}:
\begin{eqnarray}
p_{A}&=&\frac{1}{2}\left(1-\frac{3}{2}\frac{\geff}{G_0}\right) +\frac{1}{4}\sqrt{9\left(\frac{\geff}{G_0}\right)^2+12\frac{\geff}{G_0}(3-2\zeta)-20}
\label{p_A}
\end{eqnarray}
In parameterization B the exponent is different:
\begin{equation}
p_B=-\frac{1}{4}+\frac{1}{4}\sqrt{1+24\frac{\geff}{G_0}(1-\zeta)}
\label{p_B}
\end{equation}
One can see that in the GR limit ($\zeta=0$, $\geff=G_0$) we recover $p_A=p_B=1$, in agreement with the standard result for an Einstein-de Sitter universe. Equation (\ref{p_A}) implies that in a universe with \mbox{$\zeta=0$} matter perturbations fail to grow during the matter-dominated epoch for values of $\geff/G_0 < 0.5$. Hence we should expect this region of parameter space to be disfavoured by the matter power spectrum. The analogous boundaries in parameterization B will be different, which may explain the different shapes of the fall-offs of the distributions in figure 7. However, we ought to remember that  equations (\ref{p_A}) and (\ref{p_B}) apply only in a pure Einstein-de Sitter setting. The boundaries in parameter space that they imply may not be obeyed rigidly in the real universe, due to growth during the radiation- and $\Lambda$-dominated eras.

For small deviations from GR $p_a\approx p_b$.  This is because in EdS universes $\dot\Phi=0$ so the slip equations (\ref{slip}) and (\ref{phenom_slip}) are the same.  This does not apply when $\Lambda$ becomes important.

\begin{figure}[htbp]
\begin{center}
\includegraphics[width=9cm]{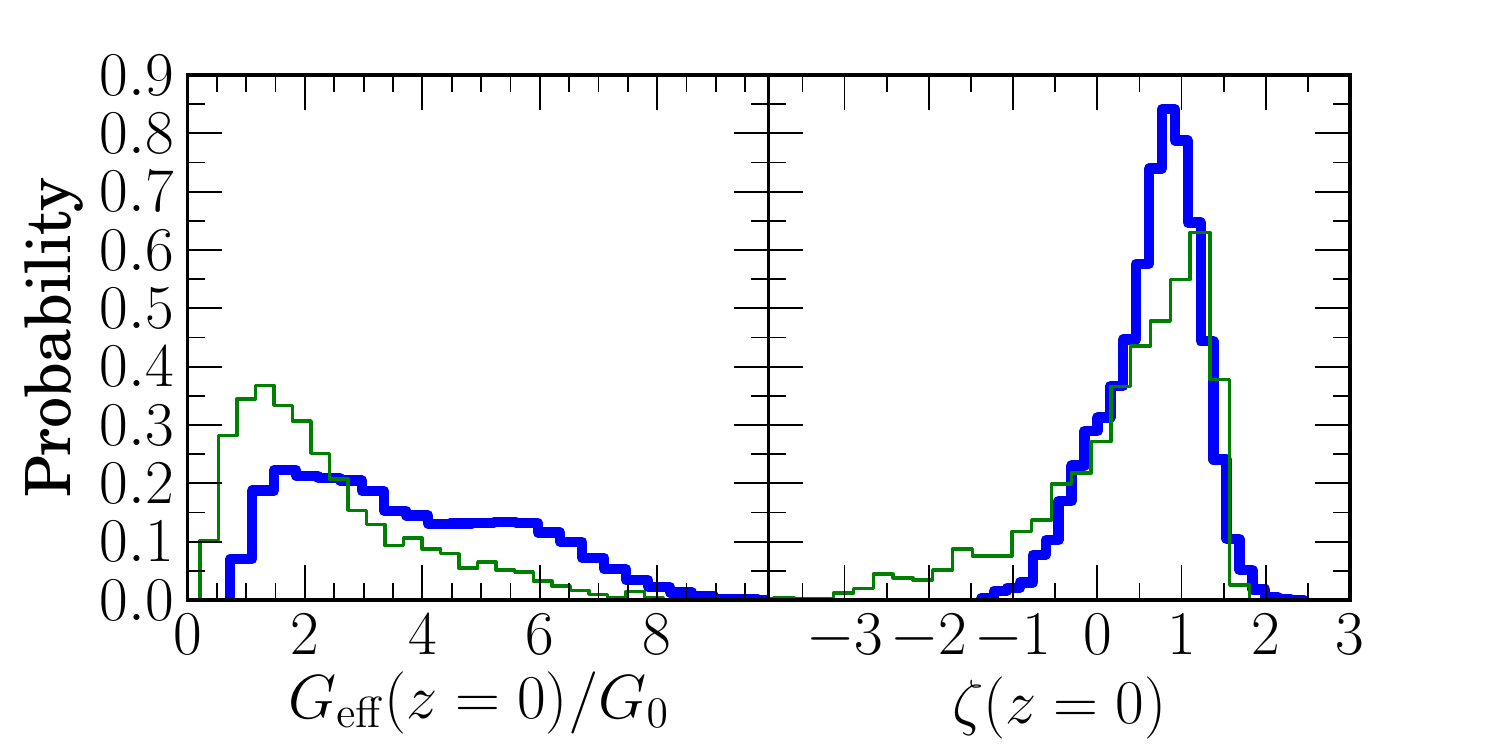}
\caption{Likelihoods of the effective gravitational and slip parameters at $z=0$, for parameterization A (thick blue) and B (thin green).  These plots are marginalized forms of figure \ref{2d histogram}.  The distribution means are significantly shifted in each case, and the tails even more so.}
\label{G0 zeta0 histogram}
\end{center}
\end{figure}

It is in the tails of the distributions where the difference between the two parameterizations can be most stark -- this will be particularly important when we are trying to decide if GR is threatened by evidence of deviations.  The histograms in figure \ref{G0 zeta0 histogram} demonstrate this most clearly.  In addition, the spectral index is much less constrained in model A than in either GR or model B, because in that model we can more freely cary $\geff$ at low redshift; so the matter power spectrum amplitude is not as constraining.

\section{Discussion}
\label{discussion}

The first, and uncontroversial, issue we have noted here is the significant effect that using insufficiently free functions has on the constraints one obtains on modifications to the slip and Poisson relations.  We now argue that there are two reasonable ways to get around this problem.  One is to embrace the constraints, but to make them correspond to some regime in theory space that we wish to model.  This is the approach that we are working towards when developing the parameterization introduced in \cite{skordis}.  The alternative is to evade the constraints as far as is possible by making the functions completely free in both $k$ and $\eta$.  This numerically challenging approach is the one taken in \cite{pca1,pca2}, where the data itself is given the freedom to choose the parameterization using a Principal Component Analysis (PCA) method.

The second argument we have advanced in this brief report is as follows. Different parameterizations -- even at the level of where the free functions are placed analytically -- can lead to radically different effects on the gravitational potentials $\Phi$ and $\Psi$, and
their effects on cosmological observables.  We have focused on the the ISW effect, which gives one of the most important constraints on parameterized approaches. In that case significant changes to the ISW plateau can be absent due to cancellations in these theories, even in the case of very large modifications. The details of this cancellation depend on which terms are included in the extension to the Poisson and slip equations. Given that the goal of these parameterized frameworks is to model and detect realistic deviations from GR, it makes sense to consider modifications to the equations that reflect the reasonable regions of theory space as well as possible, as advocated in \cite{baker}. The differences highlighted in this paper between the parameterizations underline the delicacy of this question.

Numerical approaches with very free functions can generate the same effect by numerically finding the same cancelling solutions, but models with an explicit parameterization will not in general do the same unless they are chosen for that specific purpose.

In the analysis we have undertaken here the differences between the parameterizations are smaller than the error bars on either of them. This is a temporary situation -- the data will soon improve to the point where the differences are significant.  In the longer term as they improve further they will clearly pick out which part of theory space is correct regardless of which parameterization is used, but in the intermediate phase before then the choice will matter.  Furthermore, we have focused on the ISW in the CMB but the arguments follow through to other cosmological observables, such as weak lensing, redshift space distortions and any other probe of the gravitational potentials.

It is often argued that the purpose of these methods is not to model alternative theories, but simply to detect any deviation from GR+$\Lambda$CDM in as simple a way as possible \cite{Pogosian_parameterization, Hojjati_Pogosian}.  But, as we have shown in this paper, two different parameterizations can lead to different constraints -- calling into question the interpretation of any individual detection if not properly put into the context of the class of theories that are being considered. 

Furthermore, the best way to maximize the ability of a method to find a signal like this is to model as closely as possible its expected characteristics, as in matched filter methods in signal analysis.

In future work we will be extending parameterization A to model additional field components and modified background evolution terms.  We will also be applying it to other data sets like lensing and cross-correlations which react differently to metric potentials.  An inclusive approach which carefully includes as many phenomena as possible in a model offers the best chance of detecting any small deviations from GR in the upcoming era of high-precision cosmic structure data.
\\
\emph{Acknowledgements}
We gratefully acknowledge helpful discussions with Tom Zlosnik, Levon Pogosian, Kazuya Koyama and Gongbo Zhao.  This work was supported by STFC, the ERC, the Oxford Martin School, and the Beecroft Institute of Particle Astrophysics and Cosmology.
CS is supported by a Royal Society University Research fellowship.

\bibliographystyle{JHEP}
\bibliography{paper}

\end{document}